\documentclass[a4paper]{raa}
\usepackage{graphicx,times}
\usepackage{natbib}
\usepackage{amssymb,amsmath}
\usepackage{ulem}
\usepackage{color}

\newcommand{\Ha}{\mbox{${\mathrm H\alpha}$}}
\newcommand{\Hb}{\mbox{${\mathrm H\beta}$}}

\begin{document}

    \title{LAMOST Medium-Resolution Spectral Survey of 
    Galactic Nebula (LAMOST MRS-N): An overview of Scientific goals and Survey plan}

 	\volnopage{ {\bf 2012} Vol.\ {\bf X} No. {\bf XX}, 000--000}
   	\setcounter{page}{1}

	\author{Chao-Jian Wu\inst{1,3}, Hong Wu\inst{1,3}, 
			Wei Zhang\inst{1,3}, Juan-Juan Ren\inst{2,3}, 
			Jian-Jun Chen\inst{1,3}, Chih-Hao Hsia\inst{4}, 
			Yu-Zhong Wu\inst{1,3}, Hui Zhu\inst{3}, 
			Bin Li\inst{5,6}, Yong-Hui Hou\inst{7,8}, 
			Jun-Lin Wang\inst{4}, Shuo-Ran Yu\inst{4}, 
			LAMOST MRS Collaboration}

	\institute{CAS Key Laboratory of Optical Astronomy, National Astronomical Observatories, Chinese Academy of Sciences, Beijing 100101, China; {\it chjwu@bao.ac.cn}\\
	\and
				CAS Key Laboratory of Space Astronomy and Technology, National Astronomical Observatories, Chinese Academy of Sciences, Beijing 100101, China\\ 
	\and
				National Astronomical Observatories, Chinese Academy of Sciences, 20A Datun Road, Chaoyang District, Beijing 100101, China\\
	\and
				State Key Laboratory of Lunar and Planetary Sciences, Macau University of Science and Technology, Taipa, Macau, China \\
	\and
				Purple Mountain Observatory, Chinese Academy of Sciences, Nanjing 210008, People's Republic of China\\
	\and
				University of Science and Technology of China, Hefei 230026, People's Republic of China\\		
	\and
				Nanjing Institute of Astronomical Optics, \& Technology, National Astronomical Observatories, Chinese Academy of Sciences, Nanjing 210042, China \\
    \and
                School of Astronomy and Space Science, University of Chinese Academy of Sciences \\
	\vs \no
	{\small Received ---; accepted ---}
	}

\abstract{
    Since Sep. 2018, LAMOST started the medium-resolution (R $\sim 7500$) spectral survey (MRS). We proposed the spectral survey of Galactic nebulae, including \hii\ regions, HH objects, supernova remnants, planetary nebulae and the special stars with MRS (LAMOST MRS-N). LAMOST MRS-N covers about 1700 square degrees of the northern Galactic plane within $40^\circ < l < 215^\circ$ and $-5^\circ < b < 5^\circ$. In this 5-year survey, we plan to observe about 500 thousand nebulae spectra. According to the commissioning observations, the nebulae spectra can provide precise radial velocity with uncertainty less than 1 km s$^{-1}$. These high-precision spectral data are of great significance to our understanding of star formation and evolution.
}

\authorrunning{Ch-J Wu et al.}
\titlerunning{LAMOST MRS Nebula Survey}
\maketitle

\section{Introduction}\label{s:intro}

%As a link between the study of stars and cosmological scales, t
The Milky Way is a unique laboratory for studying the formation of stars and the structure of the universe. According to the Hubble classification, the Milky Way is a typical barred spiral galaxy, which consists of disk, nuclear sphere and halo \citep{2002SSRv..100..129G}. The Galactic plane (GP) contains majority of baryonic matter in the Milky Way, and it is the main region of star formation and evolution. A large number of stars are born in the nebulae of the GP. The study of nebulae is of great significance to understand the formation and evolution of stars and the feedback of interstellar matter. Through the spectral observation of the nebulae on the GP, we can obtain the basic physical parameters, including distance, mass, velocity, electronic temperature and density, abundance of many chemical elements, and other physical parameters of the ionized gas around the star, we can also study the physical properties of the accretion disk, accretion process of the protoplanetary disk and dynamic information. Moreover, the physical mechanism and evolution process of interstellar medium (ISM) in the early and late stages of star formation can also be studied. However, there is not yet a complete spectral survey of emission nebulae on the GP. Large Sky Area Multi-Object fiber Spectroscopic Telescope (LAMOST), similar to Integral Field Unit (IFU) type structure, provides us an opportunity to make spectral observations for a large number of Galactic nebulae. 
%So, it is necessary to carry out a spectral survey of nebulae on the Galactic plane, the composition of the dust around the star and so on. 

The LAMOST, also known as Guoshoujing Telescope, is a 4-meter effective aperture quasi-meridian reflecting Schmidt telescope, which is equipped with 4000 fibers on its $5^\circ$ diameter field of view focal plane. It is the first large astronomical scientific device in China \citep{cui2012, zhao2012, 1996ApOpt..35.5155W, 2004ChJAA...4....1S}. Until 2017, LAMOST has completed its pilot and first 5 years low spectral resolution (R $\sim$ 1800) regular sky survey and achieved great success. During 2017, another new 16 medium spectral resolution (R $\sim$ 7500) spectrographs were installed on LAMOST \citep{liu2020}. The two cameras (blue and red) of each spectrograph can cover the wavelength range 4950\AA\ - 5350\AA\ and 6300\AA\ - 6800\AA, respectively \citep{wu2020, liu2020}. The red part of the spectra can well cover \Ha, [\ion{N}{ii}]$\lambda \lambda$6548,6584 and [\ion{S}{ii}]$~\lambda\lambda$6717,6731 emission lines, while the blue part can well cover [\ion{O}{iii}]$~\lambda\lambda$4959,5007 emission line. The six emission lines are all crucial for the Galactic nebulae. 

Since Sep. 2018, LAMOST started the second stage survey (the Medium-Resolution Spectral Survey, MRS) program. We proposed the spectral survey of Galactic nebulae with LAMOST MRS (hereafter: MRS-N). In addition to MRS-N, another several scientific goals of MRS are Galactic archaeology, stellar physics, star clusters, star formation, exoplanet and their host stars etc.

The structures of this paper are as follow: Section \ref{s:Sci} describes the proposed scientific goals. Section \ref{s:SP} provides a detailed report of the survey strategy, input catalog, progress and preliminary results. And Section \ref{s:summary} gives a brief summary.

\section{Science}
\label{s:Sci}
The main scientific goals of our MRS-N project focus on four-type nebulae including \hii\ regions, Herbig-Haro objects, supernovae remnants and planetary nebulae. We aim to obtain the three-dimension (3D) properties of these nebulae such as radial velocities, velocity dispersions, reddening and chemical compositions, which can help us to study  the environments surrounding star formation regions and evolved stars and to understand the feedback of star formation on the dynamics and energetics of Galactic interstellar medium (ISM). Compared to the low-resolution spectra, the medium resolution spectra can provide more accurate velocity measurements, and are more efficient to separate different components along the same line-of-sight of the nebulae based on their various velocity distributions. 

Besides nebulae, some individual objects such as Be stars \citep{wu2020}, F and OB stars, luminous blue variables (LBVs), red supergiants (RSGs), and Wolf-Rayet stars are also the scientific goals of our MRS-N project.

\subsection{\hii\ regions}

\hii\ regions are the diffuse nebulae with amorphous photoionized by ultraviolet radiations emitted from young massive stars. As the radiation sources are short-lived stars, and most of the gas are swept away by strong radiation pressure, the \hii\ regions will fade away after a few million years \citep{1988AJ.....95..720K}. As the distribution of these stars and the surrounding gas are irregular, in most situation, the shape of the \hii\ regions is far from simple Stro\"mgren sphere, instead, they often appear clumpy and filamentary \citep{2011MNRAS.413.2242M}.  From the optical spectra, we can easily identify strong hydrogen emission lines (such as \Ha, \Hb) and fine-structure lines (such as [\ion{O}{ii}]$~\lambda\lambda$3726,3729, [\ion{O}{iii}]$~\lambda\lambda$4959,5007, [\ion{N}{ii}]$~\lambda\lambda$6548,6584,  [\ion{S}{ii}] $~\lambda\lambda$6717,6731), which are usually used to derive current star formation rate, internal reddening, metallicities, the electron temperature and densities of the nebulae \citep{2017PASP..129d3001P}. So the \hii\ regions are natural laboratories for studying star-forming actives and interactions between the massive stars and surrounding ISM \citep{2018ApJ...853..151M}.

The spatial variation of chemical abundances across the galactic disc is an observational constraint for chemical evolution models. Galactic abundance gradients in the ISM were described by \cite{Searle1971} in a survey of \hii\ regions in six Sc galaxies. \cite{Shaver1983} firstly showed that the abundances were not distributed homogenously throughout the disc of the Milky Way. Since then radial abundance gradients probed by \hii\ regions have been studied extensively \citep{Simon1995, Afflerbach1997, Deharveng2000, Esteban2005, Rudolph2006, Balser2011, 2017A&A...597A..84F, Esteban2017}, and the results showed that the abundances were decreasing with Galactocentric distances, indicating that the inside-out models of galaxy formation are valid to explain the chemical composition of the Milky Way \citep{1997ApJ...477..765C, 2001ApJ...554.1044C}.

Whether the slope is uniform along the entire disc is still an open question. Some studies found there is a flatter radial abundance gradient at the outer disc \citep{1991ApJ...366..107F, 1996MNRAS.280..720V, 2013MNRAS.433..382E, 2014MNRAS.444.3301K, 2017A&A...597A..84F}, while some others did not \citep{Deharveng2000, Rudolph2006, Esteban2017}. Most recently, using \hii\ regions in the Galactic anti-center area from LAMOST DR5 and several \hii\ regions from literatures, \cite{Wang2018} determined the slopes of oxygen abundance gradient of inner and outer parts, and claimed the absence of flattening at the outer disc.

The Galactic plane survey of LAMOST-MRS-N is in the coverage of the Isaac Newton Telescope Photometric H-Alpha Survey of the Northern Galactic Plane (IPHAS) \citep{2005MNRAS.362..753D, 2014MNRAS.444.3230B}. By assigning the fibers to diffuse regions of bright H$\alpha$ emission, the LAMOST-MRS-N will obtain large numbers of optical spectra of \hii\ regions on the Galactic disc, which allow us to study the radial abundance gradient and to re-check whether there is evidence of flattening at the outer disc. By comparing the results with chemical evolution models, our understanding of formation and evolution of the Milky Way will be enchanced.

One important intrinsic characteristic of giant \hii\ regions (diameter $>$ 100 pc and H$\alpha$ luminosity $\mathrm{\sim 10^{39}~erg~s^{-1}}$ ) is that they present supersonic velocity dispersions ($\mathrm{\sigma~\ge12.85~km~s^{-1}}$ ) \citep{1970ApJ...161...33S, 2000AJ....120..752F}. There are mainly two kinds of possible mechanisms to explain the supersonic line widths, one is gravitational models, in which the Gaussian line profiles are assumed to be related to virialized motions within the \hii\ regions \citep{1981MNRAS.195..839T}, another one is wind-driven models, in which the emission line profiles are broadened by mechanical energy injected by massive stars \citep{1979A&A....73..132D, 1996ApJ...456..264T}.

\subsection{Herbig-Haro objects} \label{mrs-n:hh}
Herbig-Haro objects (HH objects, \citep{Ambartsumian1954}) are formed when narrow jets of partially ionized gas ejected by stars collide with nearby gas and dust at speeds of several hundred kilometres per second. There are more 1000 individual objects have been discovered\citep{2010A&A...511A..24D}. They are small-scale shock regions intimately associated with star forming regions \citep{Schwartz1983} and can trace supersonic shock waves \citep{Bally2016}, often found in large groups, and some can be seen around a single star, aligned with its rotational axis. 

HH objects show strong hydrogen recombination lines and various forbidden lines, in particular [\ion{S}{ii}] and [\ion{O}{ii}], but no detectable optical continuum emission. Their nature and origin remained a puzzle over quite some time, although it was clear from the beginning that they were somehow connected to star formation. A major step towards an understanding of HH objects came with the suggestion that their spectral properties might arise in gas that is shock excited by supersonic winds from the young stars \citep{Schwartz1975}. The next crucial step was the discovery of the high proper motions in HH objects indicative of space motions of several hundred $\mathrm{km\ s^{-1}}$ \citep{Cudworth1979}. Another observation led to the still widely accepted basic picture of what the majority of HH objects are: most HH objects were not independent entities (like bullets), but shock fronts in continuous, well collimated jets driven by young stellar objects \citep{Mundt1985, Mundt1987}. 

The common methods for studying HH objects are narrow band photometry and spectral observation. While the narrow band images can provide the information of only one specific band each time. To study HH objects in detail, many narrow band filters are needed, at the same time, more observation time will be taken. While the spectral observations are more efficient. Fortunately, LAMOST-MRS-N provides the opportunity to study the most of known HH objects in detail and identify more HH objects candidates.

\subsection{Supernovae Remnants}\label{sec:snr}

Supernova Remnants (SNRs) are the material ejected in an explosion of supernovae, which can continue their life through the interactions with the surrounding interstellar medium (ISM) for thousands or even up to a million years. SNRs are also the important sources of the injection of energy and heavy elements into the ISM, and are believed to be the possible sources of the Galactic cosmic rays. Moreover, SNRs probe the immediate surroundings of supernovae, shaped by their progenitors. SNRs are generally characterized by the interaction of supernova ejecta with the surrounding ISM. The main features of SNRs are strong shock waves, amplified magnetic field, ultra-relativistic particles (cosmic rays) generation and associated synchrotrion radiation \citep{2017POBeo..97....1A}. Currently, 294 Galactic SNRs \citep{2014IAUS..296..188G} have been discovered. A great majority ($\sim$\,79\%) of the known Galactic SNRs are believed to be relatively old with ages greater than 10$^3$ years and radii larger than $\sim$\,5\,pc, and are evolved objects in either the adiabatic or the early radiative stages of their evolutionary development \citep{1972ARA&A..10..129W}. These old and evolved SNRs generally show characteristic shell structures \citep{1985ApJ...292...29F}. About 30\% of Galactic SNRs are known to have optical emission associated with their nonthermal radio emission. For old remnants, the optical emission arises from the cooling of shocked interstellar cloud material following the passage of the remnant's blast waves as they expand outward into the ambient medium.

Optical spectra of SNRs usually show strong emission including that of H$\alpha$, [\ion{O}{ii}]$\lambda \lambda$ 3726,3729, [\ion{O}{iii}]$\lambda \lambda$ 4959,5007, [\ion{N}{ii}]$\lambda \lambda$ 6548,6584 and [\ion{S}{ii}]$\lambda \lambda$ 6717,6731. And the line intensity ratios have important physical implications. The optical spectra of SNRs have elevated [\ion{S}{ii}]/H$\alpha$ emission-line ratios, as compared to the spectra of normal \hii\ regions, and through this technique, more than 1000 optical extra-Galactic SNR candidates have been detected so far \citep{2015SerAJ.191...67V}. The [\ion{S}{ii}]/H$\alpha$ ratio can be also used to differentiate the shock-heated SNRs (ratios $>$ 0.4, but often considerably higher) and photoionized nebulae (0.4, but typically $<$ 0.2). As in typical \hii\ regions, sulfur exists mainly in the form of S III, thus the [SII]/H$\alpha$ ratio is very low. While after the shock wave from an SN explosion has propagated through the surrounding medium, and the material has cooled sufficiently, a variety of ionization states can be present, including the [\ion{S}{ii}], thus leading to the increased ratio [\ion{S}{ii}]/H$\alpha$ observed in SNRs \citep{2004ApJS..155..101B}. Generally, in \hii\ regions the material is too ionized to produce forbidden lines of [\ion{O}{i}], [\ion{N}{ii}] or [\ion{S}{ii}], while in SNRs, both the models of \cite{2008ApJ...683..773A} and observations show optical spectra containing these forbidden lines (\cite{2016hst..prop14638L}). Furthermore, the [\ion{S}{ii}]6717/[\ion{S}{ii}]6731 ratio is electron density sensitive, thus can be used to obtain the electron density. The H$\alpha$/[\ion{N}{ii}]6584 ratio is a widely used tool for investigating nitrogen-to-hydrogen abundance variations among SNRs. Thus it is important to obtain complete sampling of the optical spectra covering these emission in SNRs.

Although several large and evolved SNRs have been well studied morphologically in the optical, only a limited amount of spectroscopic data are available for those faint optical remnants, especially for those with large angular sizes. While many SNRs are very large, where there are 10 large SNRs with angular sizes greater than 2$^\circ$ \citep{2015ASSP...43.....O}. It would be extremely time consuming to obtain spectra covering such a large spatial extent using regular telescopes. The wide field of view, multi-fiber survey telescope LAMOST thus serves as a perfect tool for mapping the details of those extremely extended SNRs. \cite{2018RAA....18..111R} has tried to map a large SNR (the S147 with angular diameter $\sim$ 200 arcmin) with the LAMOST low-resolution (R$\sim$1800) spectra, but due to the low-resolution, the details can not be mapped clearly. With this increased higher resolution of MRS-N, more detailed emission features and kinematics properties can be mapped for SNRs, thus leading to better understandings of the properties of SNRs.

 \subsection{Planetary nebulae}\label{sec:pn}

Planetary nebulae (PNs) are originally thought to be a member of \hii\ regions, in fact they are completely different from normal emission nebulae. PNs are the evolutionary end products of low- and intermediate-mass stars (1 $\sim$ 8 M$_\odot$). The total number of Galactic PNs is about 2860, including $\sim$1500 in the Galactic PN Catalogue of \cite{2001A&A...378..843K}, $\sim$1200 PNs listed in the Macquarie-AAO-Strasbourg H$\alpha$ Catalogue of Galactic PNs (MASH; \cite{2006MNRAS.373...79P, 2008MNRAS.384..525M}), and $\sim$160 faint extended PNs discovered by the IPHAS. Most PNs are located close to the Galactic plane and suffer larger interstellar extinction. They always show low surface brightness and can be easily mixed with background diffuse nebulae or star formation regions.

With the great help of MRS-N, a significant fraction of known PNs ($\sim$ 14$\%$; 400 PNs) will be observed for studying their 3D ionized structures based on various nebular expansion velocities. The gaseous structures of Galactic PNs and surrounding nebulae are easily separated by their various spatial expansion velocities. In addition, some foundational properties of these PNs can also be derived. The survey provides us an opportunity to study 3D ionized structures and electron density distributions of Galactic PNs, the evolution of various morphological PNs, and the chemical abundance of our galaxy. 

MRS with large spatial coverage and high sensitivity as a powerful tool to study PNs obscured in the GP. Many new PNs expected to discover through this survey, containing young, compact PNs and old, extended, faint PNs. The nature and the PN status of these nebulae can be identified using the log(H$\alpha$/[\ion{N}{ii}]) vs. log(H$\alpha$/[\ion{S}{ii}]) emission-line-ratio diagnostic diagram \citep{2014A&A...563A..63H}, which is an useful probe to distinguish PNs from \hii\ regions and supernova remnants (SNRs) with ionized gaseous nebulae. 

\section{Survey plan}
\label{s:SP}
\subsection{Strategy}

Most nebulae are located on the GP. So the survey areas of MRS-N mainly cover the northern GP. The northern GP is within the Galactic longitude range $40^\circ < l < 215^\circ$ and the latitude range $-5^\circ < b < 5^\circ$, totally about 1700 square degrees. Furthermore, four large size nebulae are selected to be observed with MRS as the specific areas in our Galaxy. The four specific areas are Westerhout 5 (\hii\ region, hereafter West), Rosettle Nebula+NGC2264 (\hii\ region, hereafter Ros), Cygnus Loop (SNRs, hereafter Cyg) and Simeis 147 (SNRs, hereafter S147), respectively. 

\begin{figure}
    \centering
    \includegraphics[width=\textwidth]{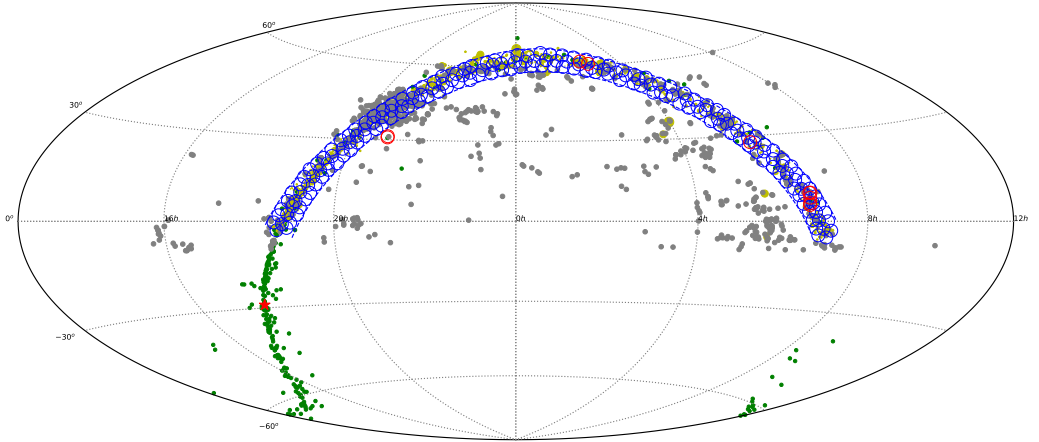}
    \caption{123 GP plates (blue circles) and four specific plates (red circles) for MRS-N in an Equatorial Aitoff projection. Most of the bright nebulae in the Milk way are marked with gray dots. The green dots represent the SNRs and the gray ones represent the \hii\ region. The red star is the location of the Galactic center.}
    \label{fig:cover}
\end{figure}

In MRS-N, we plan to use 123 plates cover the GP areas (blue circles in Figure \ref{fig:cover}). \ref{fig:cover} shows that the MRS-N coverage can cover most of \hii\ region and SNRs in the northern celestial sphere. A plate, covering about 20 $\mathrm{deg^2}$ (a circle with a diameter of $5^\circ$), will continuously take three single 1200 s (900 s for MRS-N, the exposure time of 900 s can provide enough high signal-to-noise ratio nebulae spectra) exposure and obtain the co-added spectra with total exposure time of $3 \times 1200$ s ($3 \times 900$ s for MRS-N). For each specific area, the exposure time is designed as 16 $\times$ 900 s (red circles in Figure \ref{fig:cover}). So it will take about 130 hours to finish the whole MRS-N survey. MRS-N will be assigned 30 hours per year. So we could finish the MRS-N survey within 5 years without considering other objective factor such as weather. 

\begin{figure*}
    \centering
    \includegraphics[width=0.7\textwidth]{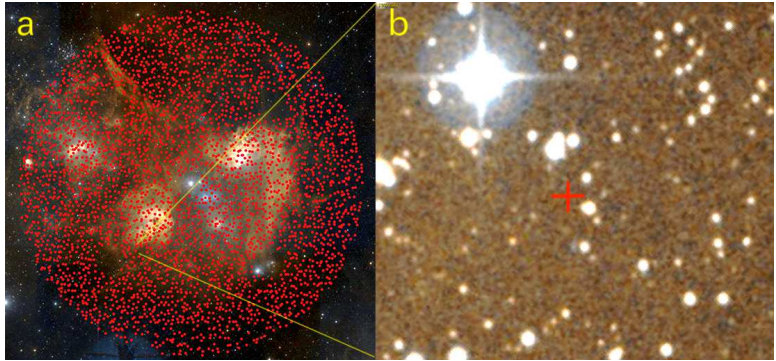}
    \caption{($a$): Fiber distribution of the commissioning observation on Mar. 4, 2018. ($b$): One fiber pointing to the nebula, its spectrum is drawn in Figure \ref{fig:abp_em}.
    \label{fig:specneb}}
\end{figure*}

\begin{figure*}
    %\centering
    \includegraphics[width=\textwidth]{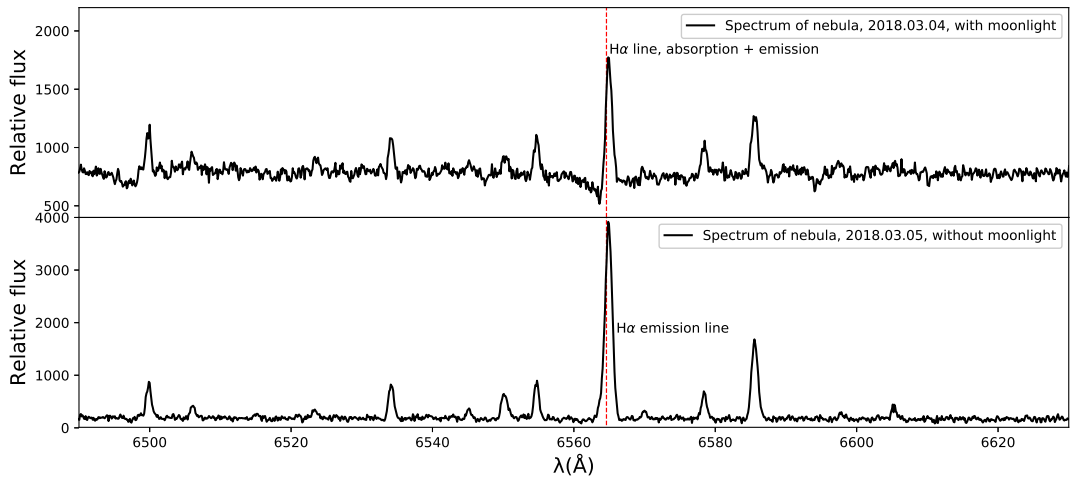}
    \caption{Two spectra of commissioning observation. The upper panel shows a nebula spectrum observed on Mar. 4, 2018, when the moon had already risen. It is clearly to see that the \Ha\ absorption and emission lines at 6563 \AA\ were mixed together. The bottom panel shows the nebula spectrum observed on Mar. 5, 2018, before the moon rose.}
    \label{fig:abp_em}
\end{figure*}

The commissioning observations were carried out on Mar. 4-5, 2018. Figure \ref{fig:specneb} (a) shows the fiber distribution. More detailed observation information are listed in Table \ref{tab:obs}. The results showed that MRS-N is seriously affected by the moonlight. The spectrum of one fiber (red plus in Figure \ref{fig:specneb} $b$) is presented in Figure \ref{fig:abp_em}. The \Ha\ absorption and emission line at 6563 \AA\ are obviously mixed together. However, the absorption lines are not visible at [\ion{N}{ii}]($\lambda \lambda$ 6548,6584) and [\ion{S}{ii}]($\lambda \lambda$ 6717,6731). In fact, the \Ha\ absorption line comes from the moonlight, to be exact, from the sunlight. On closer examination, we found that the spectrum was no longer affected when the altitude of the moon was lower than $1^\circ$. So all the observation time of MRS-N should be assigned into a time range before the moon rises or after the moon sets. This will cause the observation time of MRS-N to be very urgent, which is why we adjust the sub-exposure time to 900s. 

\begin{table}
    \centering
\caption{The commissioning observation of MRS-N}
\label{tab:obs}    
\begin{tabular}{cccc}
\hline\noalign{\smallskip}
    Date & Time (UT) & Exposure (s) & Alt of Moon  \\
\noalign{\smallskip}\hline\noalign{\smallskip}
    2018-03-04 & 13:19:32 & 900  & +$6.5^{\circ}$ \\
    2018-03-05 & 13:01:53 & 900  & -$8^{\circ}$ \\
\noalign{\smallskip}\hline
\end{tabular}
\end{table}

\subsection{Input catalog}
\label{s:inputcat}

The input catalog is very important to LAMOST, which can be used to allocate the fiber's coordinates. MRS-N mainly focus on the spectra of non-point source (such as the structure of nebulae). There is not yet a catalog containing the structure of nebulae. Therefore, the input catalogs of MRS-N should be made independently unlike other MRS projects. 

The input catalogs of four specific areas are man-made by comparing with optical images from the Digitized Sky Surey (DSS) and Narrow Band Survey (NBS). From optical images, some bright stars, some sky areas and the positions of interesting structure of nebulae are selected to compose input catalogs. 

\begin{figure}
    \centering
    \includegraphics[width=0.8\textwidth]{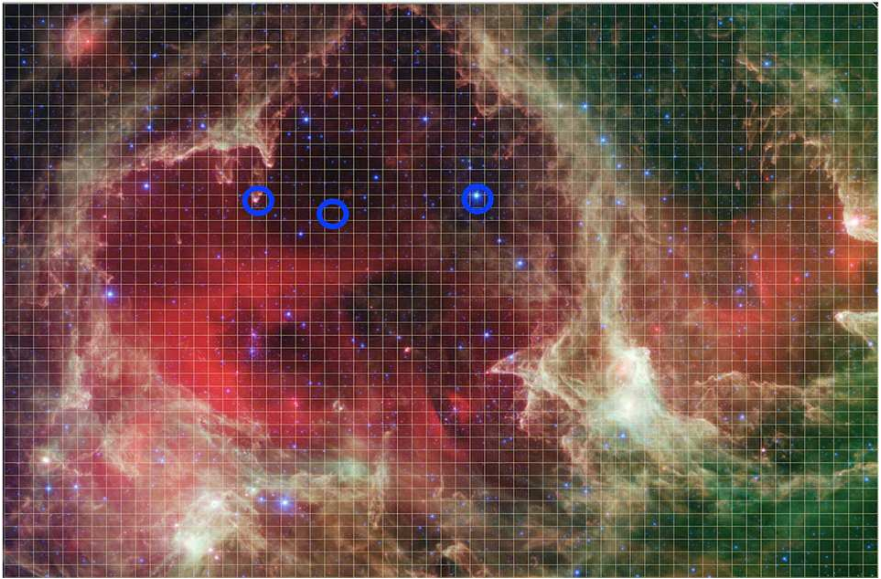}
    \caption{Creating the input catalog of MRS-N. The size of each grid is 2\arcmin $\times$ 2\arcmin. If a star within a grid is brighter than 13 magnitude (r band), then the star is selected as location of fiber (right blue circle), else if there is not a star in a grid or the star is too faint to observe, the location of fiber will be pointed to the brightest position (left and middle blue circle). The background image is Westerhout 5.}
    \label{fig:inputcat}
\end{figure}

The input catalog of GP area is very complex. We divide the northern GP area into about 1.5 million 2\arcmin $\times$ 2\arcmin grids. Each grid can be considered as the moving area of each fiber. Then we cross-match the 1.5 million grids with Hipparcos main catalog and Pan-STARRS1 (PS1). If a star within a grid is brighter than 13 mag (r band), then the star is selected as location of fiber with objtype $STAR$. By comparing with IPHAS and NBS images, if a star within a grid is bright than 18 mag and fainter than 13 mag, then the location of fiber within the same grid should avoid the star with objtype $skylight$ or $NEB$ (Nebulae). Else if there is not a star in a grid or the star is too faint to be observed, then the location of fiber should be pointed to the brightest position with objtype $NEB$ within the grid (Figure \ref{fig:inputcat}). In addition, we integrated some F stars, OB stars and emission stars from IPHAS into the input catalog of GP area. 

Eventually, we finished a complete input catalog of MRS-N with more than 1.5 million coordinates.

\subsection{Data description and Preliminary results}
\label{s:pro_res}
The data processing of MRS-N is different from LAMOST 1D pipeline \citep{2012RAA....12.1243L, 2015RAA....15.1095L}. LAMOST 1D pipeline gives the physical parameters after deducting the skylight background. However, the method of deducting skylight background is not applicable to the nebulae spectra. How to reduce the skylight in the spectra of the nebula is still under study. Therefore, MRS-N needs to have its own pipeline. The original nebulae spectra we get are the 1D spectra given by LAMOST 2D pipeline. The process includes dark and bias subtraction, flat field correction, spectral extraction, the first wavelength calibration. The MRS-N pipeline developed by MRS-N team includes deduction of cosmic rays, merging sub-exposures, fitting skylight emission line, reducing skylight background (optional), secondary wavelength calibration and parameters measurement of nebula emission line (such as radial velocity, intensity ratio of emission line, relative flux, line width, intensity and so on). More details about pipeline, data processing and data release will be presented in Wu et al. (2020 in prep). 

Due to the more complex data processing, the data release of MRS-N is one year later than that of LAMOST. According to the LAMOST data release policy, the official data release has a protection period of 18 months. That is to say, the nebulae spectra of first year (from the second half of 2018 to the first half of 2019) will be officially released in March 2021, the rest will be done in the same manner. It is important to note that the nebula spectra in LAMOST DR7 has been processed by subtracting the skylight background and the results are not up to the release standard of MRS-N. If someone want to use nebula spectra for science, they should use the data released by MRS-N.

%Since Oct. 2018, we have completed two years of MRS-N observation. The first year of observation is from Oct. 2018 to Jan. 2019. A total of 24 days were available for observation. Due to weather and other reasons, only 9 days were suitable for observation and finally 12 plates were finished. Figure \ref{fig:year} shows the first year coverage of MRS-N. Gray circles are finished plates.

%\begin{figure}
%    \centering
%    \includegraphics[width=\textwidth]{mrs_n_cover_year2.eps}
%    \caption{MRS-N coverage of two years' observation. Gray circles are the plates observed in the first year, green circles are observed in the second year.}
%    \label{fig:year}
%\end{figure}

%The second year of observation is from Nov. 2019 to Mar. 2020. A total of 18 days were available for observation. We have optimized the observation strategy comparing with the first year, then we finished 31 plates ( green circles in Figure \ref{fig:year}). Among the 31 plates, 20 plates are used to cover the Ros area, 2 plates are used to cover the West area and the left 9 are used to cover the GP area. In the second year, we finished a complete specific area, the Ros area. 

%The MRS-N data processing of first year has been completed. The cosmic ray is removed by combining three 900s spectra. 

%\subsection{Preliminary Results}
%\label{s:res}

\begin{figure}
    \centering
    \includegraphics[width=\textwidth]{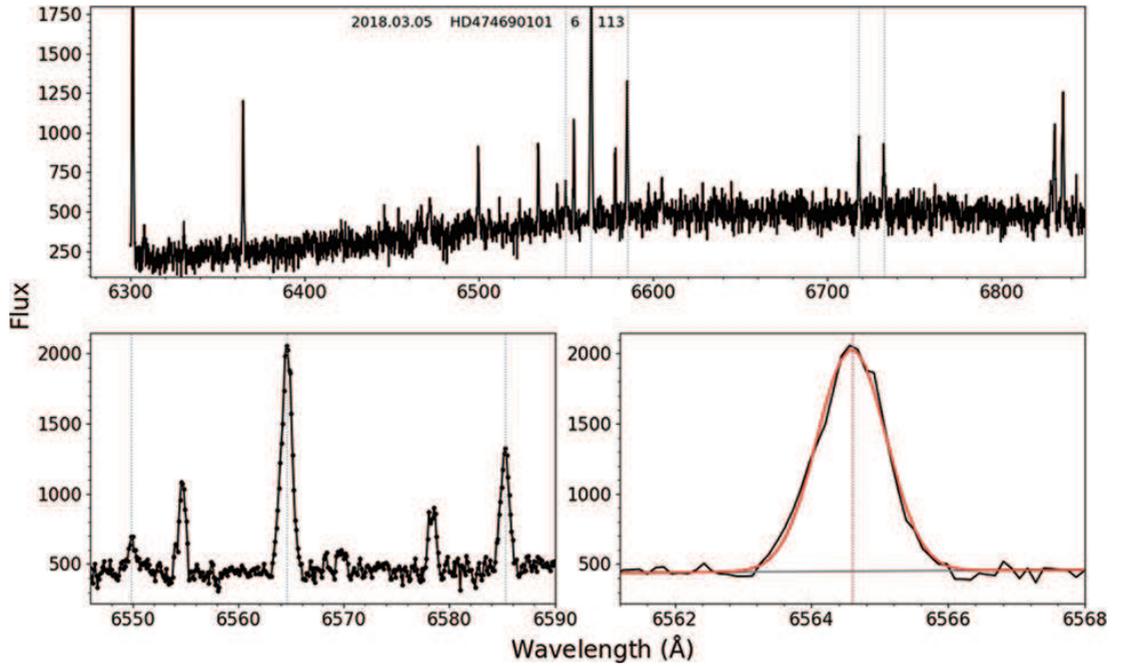}
    \caption{H$\alpha$ emission line of nebulae with single-component.}
    \label{fig:Ha1}
\end{figure}

\begin{figure}
    \centering
    \includegraphics[width=\textwidth]{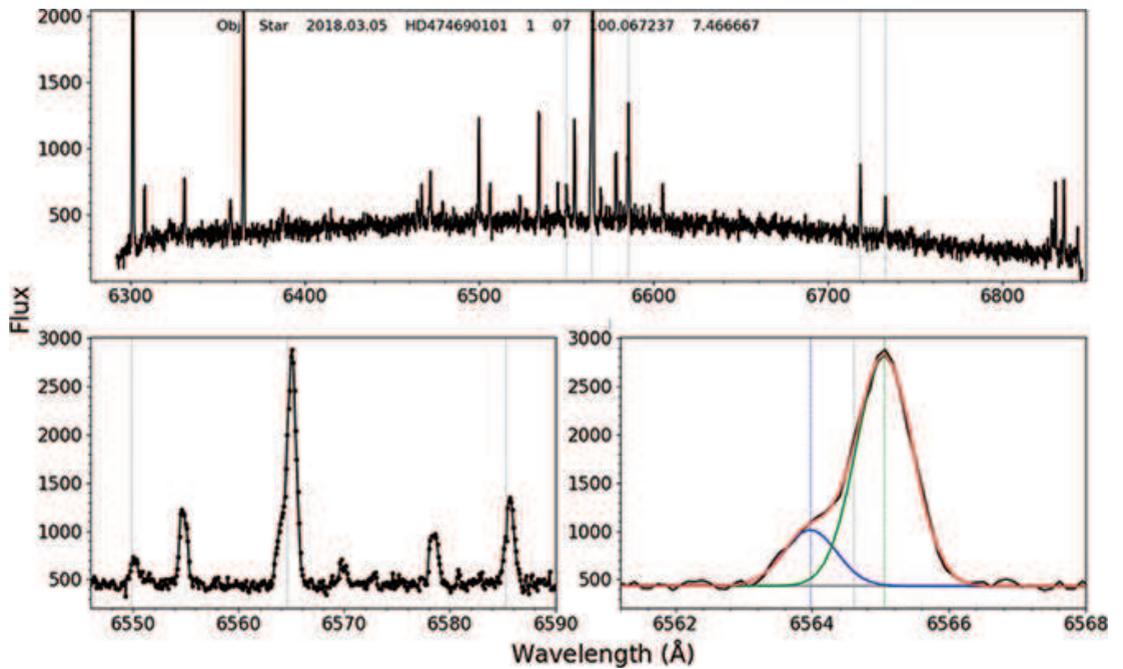}
    \caption{H$\alpha$ emission line of nebulae with double-component.}
    \label{fig:Ha2}
\end{figure}

\begin{figure}
    \centering
    \includegraphics[width=\textwidth]{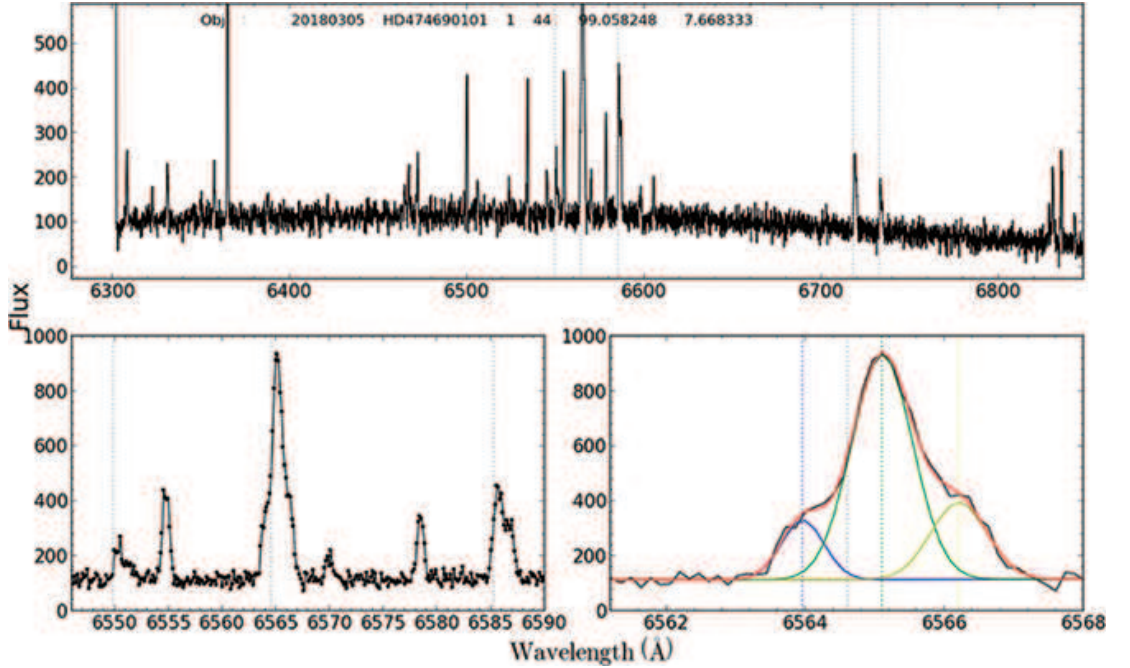}
    \caption{H$\alpha$ emission line of nebulae with multi-component.}
    \label{fig:Ha3}
\end{figure}

Figure \ref{fig:Ha1}, \ref{fig:Ha2}, \ref{fig:Ha3} are the red part nebulae spectra observed on Mar. 5, 2018, which show three various H$\alpha$ emission line composited with different components. The double and multiple components may be the result of the emissions produced by the nebulae with various expansion velocities along the line of sight. 

\vspace{30pt}

\vspace{30pt}

Ren et al. (2020, submit) investigated the precision of wavelength calibration and found that there are systematic deviations of radial velocities (RVs) from $\sim 0.1$ to 4 $\mathrm{~km~s^{-1}}$ for different plates. After by fitting 7 sky emission lines in the red part of MRS to recalibrate the wavelength, the precision of RVs is less than $\mathrm{1~km~s^{-1}}$. All the future MRS-N data will be recalibrated with this method.

\section{Summary}
\label{s:summary}
As the first large optical astronomical scientific device in China, LAMOST has achieved great success for the first stage of low-resolution spectroscopic survey. Since Sep. 2018, LAMOST started to conduct medium-resolution spectroscopic survey and will significantly increase the number of stellar and nebulae spectra with medium-resolution. As a sub-project of MRS,  we proposed  the spectral survey of Galactic nebulae with MRS. In MRS-N, the RVs, velocity dispersions, reddening and chemical compositions of most nebulae on the northern GP can be studied. Now, MRS-N has completed its two-year observation plan, which is the current largest nebulae survey over the world. In this paper, the detailed scientific goals and survey plan are presented. The first data release of MRS-N is scheduled for the third quarter of 2020 (Wu et al. 2020, in prep). 

\normalem
\begin{acknowledgements}
This project is supported by the National Natural Science Foundation of China (Grant Nos. 11733006, 11403061, 11903048, U1631131, 11973060, U1531118, 11403037, 11225316, 11173030, 11303038, Y613991N01, U1531245 and 11833006), and the Key Laboratory of Optical Astronomy, National Astronomical Observatories, Chinese Academy of Sciences, and the Key Research Program of Frontier Sciences, CAS (Grant No. QYZDY-SSW- SLH007).

C.-H. Hsia acknowledges the supports from the Science and Technology Development Fund, Macau SAR (file Nos. 119/2017/A3, 061/2017/A2 and 0007/2019/A) and Faculty Research Grants of the Macau University of Science and Technology (No. FRG-19-004-SSI).
    
Guoshoujing Telescope (the Large Sky Area Multi-Object Fiber Spectroscopic Telescope LAMOST) is a National Major Scientific Project built by the Chinese Academy of Sciences. Funding for the project has been provided by the National Development and Reform Commission. LAMOST is operated and managed by the National Astronomical Observatories, Chinese Academy of Sciences.

\end{acknowledgements}

\bibliographystyle{raa}
\bibliography{ms}

\end{document}